\documentclass[twocolumn]{aastex61}

\usepackage{color}

\def\ie{{\it i.e.,~}}

\def\kms{~{\rm km~s^{-1}}}
\def\cm3{~{\rm cm^{-3}}}

\def\muG{~{\mu\rm G}}

\received{}
\revised{}
\accepted{}

\submitjournal{The Astrophysical Journal}

\shorttitle{Effects of Alfv\'enic Drift on Diffusive Shock Acceleration}
\shortauthors{Kang \& Ryu}

\begin{document}

\title{Effects of Alfv\'enic Drift on Diffusive Shock Acceleration at Weak Cluster Shocks}

\author{Hyesung Kang}
\affiliation{Department of Earth Sciences, Pusan National University, Busan 46241, Korea}
\author{Dongsu Ryu}
\affiliation{Department of Physics, School of Natural Sciences, UNIST, Ulsan 44919, Korea}
\correspondingauthor{Dongsu Ryu}
\email{ryu@sirius.unist.ac.kr}

\begin{abstract}


Non-detection of $\gamma$-ray emission from galaxy clusters has challenged diffusive shock acceleration (DSA) of cosmic-ray (CR) protons
at weak collisionless shocks that are expected to form in the intracluster medium.
As an effort to address this problem, we here explore possible roles of Alfv\'en waves self-excited via resonant streaming instability 
during the CR acceleration at parallel shocks.
The mean drift of Alfv\'en waves may either increase or decrease the scattering center compression ratio,
{ depending on the postshock cross-helicity,}
leading to either flatter or steeper CR spectra.
We first examine such effects at planar shocks, based on the transport of Alfv\'en waves in the small amplitude limit.
For the shock parameters relevant to cluster shocks, Alfv\'enic drift flattens the CR spectrum slightly, 
resulting in a small increase of the CR acceleration efficiency, $\eta$.
We then consider two additional, physically motivated cases: (1) postshock waves are isotropized via MHD and plasma processes across the shock transition
and (2) postshock waves contain only forward waves propagating along with the flow due to a possible gradient of CR pressure behind the shock.
In these cases, Alfv\'enic drift could reduce $\eta$ by as much as a factor of 5 for weak cluster shocks.
For the canonical parameters adopted here, we suggest $\eta\sim10^{-4}-10^{-2}$ for shocks with
sonic Mach number $M_{\rm s}\approx2-3$.
The possible reduction of $\eta$ may help ease the tension between non-detection of $\gamma$-rays
from galaxy clusters and DSA predictions.

\end{abstract}


\section{Introduction}

Weak shocks with sonic Mach number typically $M_{\rm s} \lesssim$ a few are expected to form 
in the intracluster medium (ICM) during the course of hierarchical clustering of the large-scale 
structure of the Universe \citep[e.g.][]{ryu03,kang2007}. 
The presence of such shocks has been established by X-ray and radio observations of many merging clusters \citep[e.g.][]{markevitch07,brug12,brunetti14}. 
In particular, diffuse radio sources known as radio relics, located mostly in cluster outskirts,
could be explained by cosmic-ray (CR) electrons
(re-)accelerated via diffusive shock acceleration (DSA) at quasi-perpendicular shocks \citep[e.g.][]{vanweeren10, kang12, kang17}.
Although both CR electrons and protons are known to be accelerated at astrophysical shocks such as Earth's bow
shocks and supernova remnant shocks \citep[e.g.,][]{bell78, dru83,blaeic87},
the $\gamma$-ray emission from galaxy clusters, which would be a unique signature of CR protons, has not been
detected with high significance so far \citep{ackermann14,ackermann16,brunetti17}.

In galaxy clusters, diffuse $\gamma$-ray emission can arise from inelastic collisions of CR protons with thermal protons,
which produce neutral pions, followed by the decay of pions into $\gamma$-ray photons \citep[e.g.,][]{miniati01,brunetti14,brunetti17}.
Using cosmological hydrodynamic simulations, the $\gamma$-ray emission has been estimated 
by modeling the production of CR protons at cluster shocks in several studies 
\citep[e.g.,][]{ensslin2007,pinzke10, vazza16}.
In particular, \citet{vazza16} tested several different prescriptions for DSA efficiency by comparing
$\gamma$-ray flux from simulated clusters with Fermi-LAT upper limits of observed clusters.
They found that non-detection of $\gamma$-ray emission could be understood, 
only if the CR proton acceleration efficiency at weak cluster shocks is 
{  on average less than $10^{-3}$ for shocks with $M_s=2-5$.}
On the other hand, recent hybrid plasma simulations demonstrated that about $5 - 15 \%$ of the shock kinetic energy is
expected to be transferred to the CR proton energy at quasi-parallel shocks
with a wide range of Alfv\'en Mach numbers, $M_{\rm A}$, \citep{caprioli14a}.
So there seems to exist a tension between the CR proton acceleration efficiency predicted by DSA theory and
$\gamma$-ray observations of galaxy clusters.

It is well established that CR protons streaming along magnetic field lines upstream of parallel shock resonantly
excite Aflv\'en waves with wavenumber $k \sim 1/r_{\rm g}$ via two-stream instability,
where $r_g$ is the proton Larmor radius \citep{wentzel74,bell78,lucek00,schure12}.
These Aflv\'en waves are circularly polarized in the same sense as the proton gyromotion, \ie left-handed circularly polarized
when they propagate parallel to the background magnetic field.
The waves act as {\it scattering centers} that can scatter CR particles in pitch-angle both upstream and downstream of the shock, 
leading to the Fermi first order (Fermi I) acceleration at parallel shocks \citep{bell78}.

Since CRs are scattered and isotropized in the mean wave frame,
the spectral index $\Gamma$ of the CR energy spectrum, $N(E) \propto E^{-\Gamma}$, 
is determined by the convection speed of scattering centers 
in the shock rest frame, $u+u_{\rm w}$, instead of the gas flow speed, $u$ \citep{bell78}.
Here, $u_{\rm w}$ is the mean speed of scattering centers in the local fluid frame, or the speed of so-called {\it Alfv\'enic drift}.
The direction and amplitude of Alfv\'enic drift depend on the difference between the intensity of {\it forward} waves
(moving parallel to the flow) and that of {\it backward} waves (moving anti-parallel to the flow), \ie
$(\delta B^{\rm f})^2 - (\delta B^{\rm b})^2$ \citep{skill75}. 
If forward and back waves have the same intensity or if waves are completely isotropized, \ie $(\delta B^{\rm f})^2 = (\delta B^{\rm b})^2$,
then $u_{\rm w}\approx 0$.

\begin{figure}[t!]
\centering
\includegraphics[width=87mm]{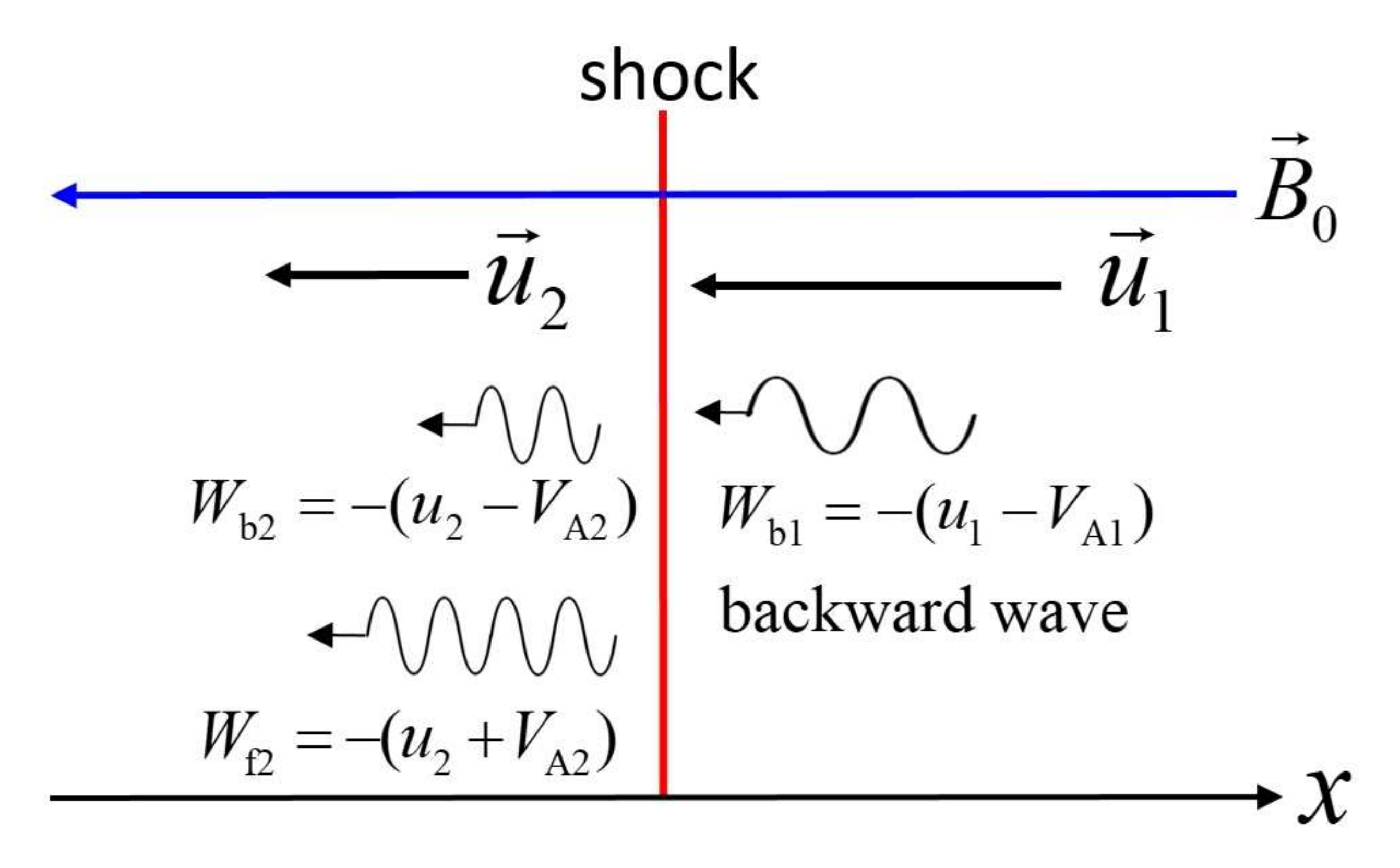}
\vskip -0.3cm
\caption{Flow velocity configuration {\it in the shock rest frame} for a 1D planar shock
with the background magnetic field parallel to the shock normal (parallel shock).
Here, the subscripts 1 and 2 are for preshock and postshock quantities, respectively.
The shock faces to the right, so the preshock flow speed is $u=-u_1$.
After upstream backward waves (moving anti-parallel to the flow in the flow rest frame) cross the shock, 
both transmitted backward waves and reflected forward waves are advected downstream.
The convection speeds of waves, $W_{\rm b1}$, $W_{\rm b2}$, and $W_{\rm f2}$, are given in the shock rest frame.}
\end{figure}

\begin{figure*}[t!]
\centering
\includegraphics[width=180mm]{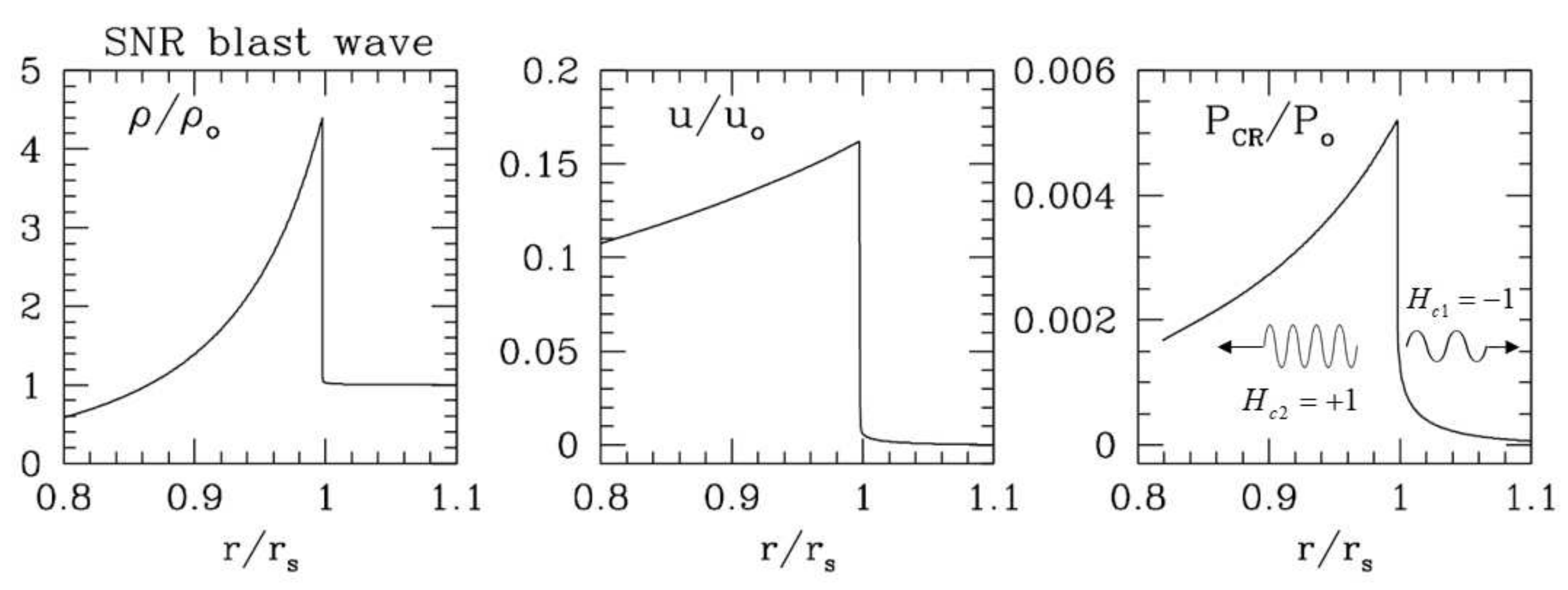}
\vskip -0.3cm
\caption{Radial profiles of the gas density, flow speed, and CR pressure of a model spherical SNR shock that expands outward.
Owing to the positive (negative) gradient of $P_{\rm CR}$, forward (backward) waves are expected to be dominant
in the postshock (preshock) region, as illustrated in this figure.
So the mean convection velocities of scattering centers point away from the shock both in the upstream and downstream
rest frames.
}
\end{figure*}

A nonresonant instability due to the electric current associated with CRs escaping upstream is also known to operate
on small wavelengths \citep{bell04,schure12}. 
The excited waves are not Alfv\'en waves, and have a circular polarization opposite to the sense of the proton gyromotion, 
\ie are right-handed circularly polarized when they propagate parallel to the background magnetic field.
This nonresonant instability is more unstable at higher $k$'s (smaller wavelengths), and the ratio of the growth rates of
non-resonant to resonant instability is roughly, $\Gamma_{\rm nonres}/\Gamma_{\rm res} \sim M_{\rm A}/30$ \citep{caprioli14b}.
In cluster outskirts where the magnetic field is observed to have $B \sim 1 \muG$ \citep[e.g.,][]{gf04},
shocks have $M_{\rm A} \lesssim 30$ (see below), so resonant instability is expected to be dominant there.
Since we here are interested in cluster shocks, we focus mainly on Alfv\'en waves excited by resonant streaming instability.

\citet{bell78} noted that resonant instability would produce mostly backward waves in the preshock region, 
because CR protons streaming upstream excite waves that move parallel to the streaming direction
(that is, travel upstream away from the shock in the upstream rest frame), 
and any forward waves pre-existing in the preshock flow would be damped due to the gradient of the CR distribution 
in the shock precursor \citep{wentzel74, skill75,lucek00}.
Then, the Alfv\'enic drift speed in the preshock region
may be approximated as $u_{\rm w1} \approx + V_{\rm A1}$, where $V_{\rm A}=B_0/\sqrt{4\pi \rho}$ is the local Alfv\'en speed.
See Figure 1 for the velocity configuration in the shock rest frame.
Hereafter, the subscripts 1 and 2 refer to the quantities in the preshock and postshock regions, respectively.

Alfv\'enic drift in the postshock region was previously considered in studies of CR acceleration 
at strong supernova remnant (SNR) shocks \citep[e.g.,][]{zp08,zp12,caprioli09,lee12,kang13jkas}.
Those studies suggested that owing to the positive gradients of the CR pressure, $P_{\rm CR}$, forward waves
(moving away from the shock toward the center of supernova explosion) could be dominant in the postshock region, 
then $u_{\rm w2} \approx - V_{\rm A2}$ (see Figure 2).

The effects of Alfv\'enic drift should be substantial, only if the Alfv\'en speed is a significant
fraction of the flow speed.
In SNR shocks, for instance, the Alfv\'en Mach number is $M_{\rm A}= u_1/V_{\rm A} \sim 20-200$,
depending on the density of the background medium, yet the Alfv\'enic drift effects could be appreciable
\citep[e.g.,][]{caprioli09,kang13jkas}.
For the ICM in cluster outskirts, the sound and Alfv\'en speeds are given as
$c_{\rm s}\approx 1.14\times 10^3\kms (k_{\rm B}T/5~{\rm keV})^{1/2}$ and
$V_{\rm A}\approx 184 \kms (B/1 \muG) (n_{\rm H}/10^{-4} \cm3)^{-1/2}$, respectively, so
\begin{equation}
\beta \equiv \left({c_{\rm s}\over V_{\rm A}} \right)^2 \approx
40 \left({n_{\rm H}\over 10^{-4} \cm3}\right) \left({k_{\rm B}T\over 5.2~{\rm keV}}\right) \left({B\over 1 \muG}\right)^{-2},
\label{eqbeta}
\end{equation}
where $k_{\rm B}$ is the Boltzmann constant.
For $M_{\rm s} \approx 2-3$, the Alfv\'en Mach number of cluster shocks ranges
$M_{\rm A}=\sqrt{\beta} M_{\rm s} \approx 13-19$, which is smaller than that of SNR shocks. 
Thus, we expect that the Alfv\'enic drift could have non-negligible effects on DSA at cluster shocks.
Note that this definition of $\beta$ differs from the usual plasma beta by a factor of 1.2 for 
the gas adiabatic index $\gamma = 5/3$; the plasma beta of the ICM has been estimated to be $\sim 50 - 100$
\citep[e.g.,][]{ryu08,porter2015}.

The transmission and reflection of upstream Alfv\'en waves at shocks can be calculated by solving
conservation equations across the shock transition \citep[e.g.,][]{cs92, vs98, vs99, caprioli09}.
\citet{vs98}, for instance, used the conservation of mass flux, transverse momentum, and tangential
electric field to calculate them, in the small wave amplitude limit ($b\equiv \delta B/B \ll 1$)
in the one-dimensional (1D) plane-parallel geometry.
They showed that after purely backward waves cross the shock, forward waves are also generated in the postshock region.
\citet{vs99} (hereafter VS99) extended the work by including the pressure and energy flux of waves across the shock.
The transmission and reflection of Alfv\'en waves and so the ensuing CR spectrum
are governed by $M_{\rm A}$, $\beta$, $b$, and the properties of upstream waves.
For certain shock parameters, the effective compression ratio, $r_{\rm sc}$,
which is defined as the velocity jump of scattering centers (see Section 3),
can be even larger than the gas compression ratio, $r$, leading to a flatter CR energy spectrum.

In this paper, we first estimate the effects of Alfv\'enic drift on the DSA of protons for 1D planar shocks
in high beta ($\beta \ge 1$) plasmas, with the transport of Alfv\'en waves across the shock transition described in VS99.
We then consider two other cases, which are physically motivated:
(1) postshock waves are isotropized, \ie $u_{\rm w2} \approx 0$,
and (2) forward waves are dominant in the postshock region, \ie $u_{\rm w2} \approx - V_{\rm A2}$.
We examine the Alfv\'enic drift effects in these cases too.

In the next section, the transmission and reflection of upstream Alfv\'en waves at 1D planar shocks 
are described.
In Section 3, the effects of the drift of Alfv\'en waves
are discussed with the power-law CR proton spectrum in the test-particle limit.
A brief summary including implications of our results at weak cluster shocks is given in Section 4.

\section{Transmission and Reflection of Alfv\'en Waves at Shocks}

\begin{figure*}[t!]
\centering
\includegraphics[trim=5mm 80mm 5mm 5mm, clip, width=180mm]{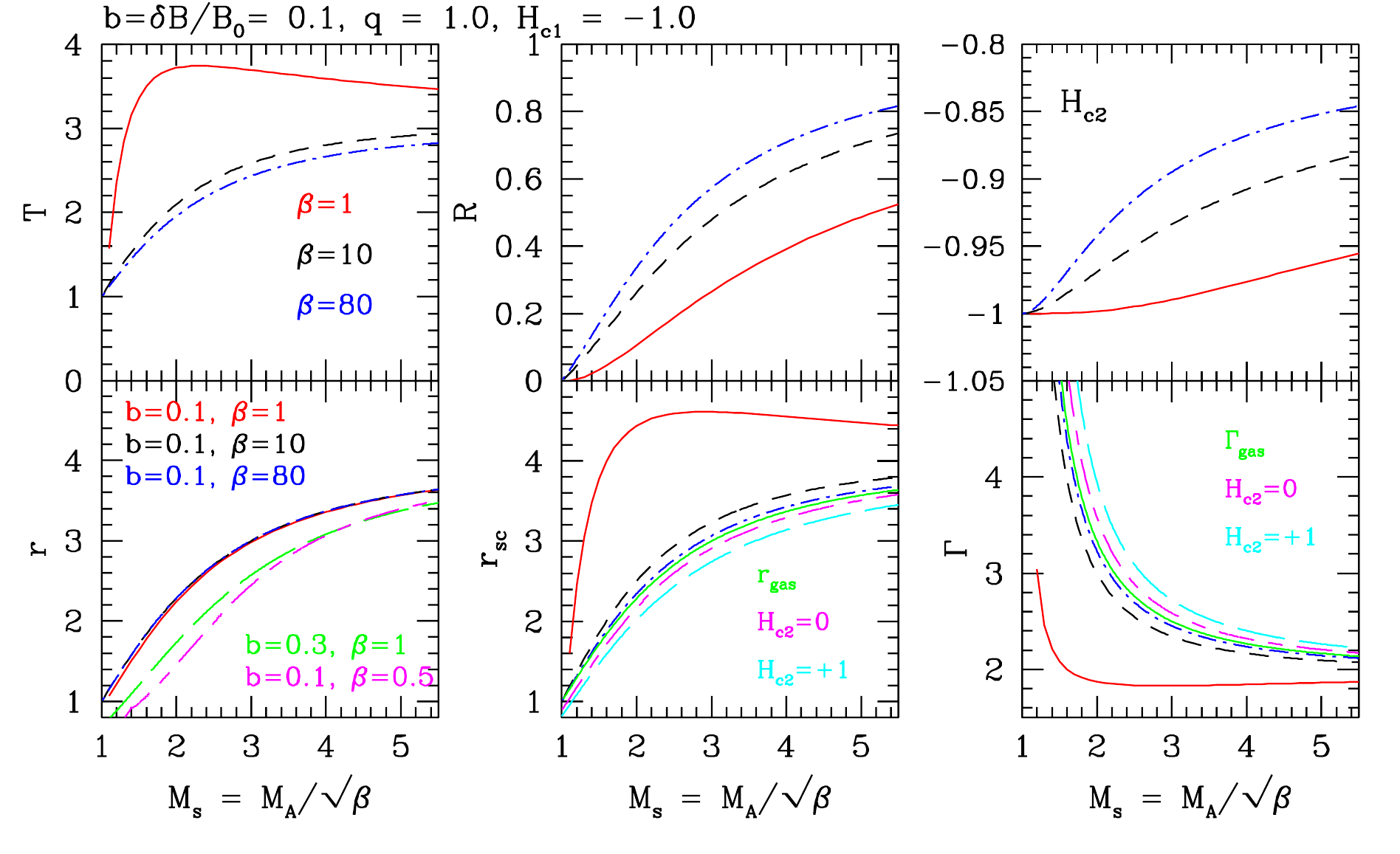}
\vskip -0.5cm
\caption{Top: Transmission and reflection coefficients, $T$ and $R$, and downstream cross-helicity, $H_{c2}$,
as functions of $M_{\rm s}$, for three cases with different $\beta$'s.
Bottom: Gas compression ratio, $r$, scattering center compression ratio, $r_{\rm sc}$, 
and CR spectral index, $\Gamma$, for the same cases.  
Here, we assume that the upstream cross-helicity is $H_{\rm c1}=-1.0$ (backward waves only) 
and the turbulence power spectrum is specified with the slope, $q=1.0$, and $b=0.1$.
{  In the panel for $r$, two additional cases are shown, the one with $b=0.3$ $\&$ $\beta=1$ by the green line,
and that with $b=0.1$ $\&$ $\beta=0.5$ by the magenta line.}
In the panels for $r_{\rm sc}$ and  $\Gamma$, the magenta lines are for the model with 
$H_{c2}=0$ (isotropic waves),
while the cyan lines are for the model with $H_{c2}=+1$ (forward waves only).
The green solid lines show $r_{\rm gas}$ and $\Gamma_{\rm gas} = (r_{\rm gas}+2)/(r_{\rm gas}-1)$
for gasdynamics shocks without Alfv\'enic drift.
}
\end{figure*}

VS99 derived necessary jump conditions for the transport of Alfv\'en waves across parallel shocks,
whose configuration is illustrated in Figure 1.
We here repeat some of them to make this paper self-contained.
The shock moves to the right, so the preshock and postshock flow speeds in the shock rest frame are
${\bold u_1}= - u_1 \hat x$ and ${\bold u_2}= - u_2 \hat x$, respectively.
The background magnetic field is given as ${\bold B_0}= - B_0 \hat x$.
CR protons streaming upstream along ${\bold B_0}$ excite backward waves that travel anti-parallel
to the background flow in the local fluid frame.
The shock amplifies the incoming backward waves and also generates forward waves in the postshock region.
The convection speed of backward waves is $W_{\rm b1,2} = -(u_{1,2}-V_{\rm A1,2}) <0$ (to the left) both upstream and downstream
of a parallel shock for the high beta plasmas with $\beta \ge 1$ considered here.

We consider nondispersive, circularly-polarized Alfv\'en waves with small amplitudes ($b\equiv \delta B/B \ll 1$),
propagating along the mean background magnetic field, ${\bold B_0}$, at 1D planar shocks.
{  Note that the formulae below do not differentiate the handedness of wave polarization, 
since the conservation equations do not depend on it.}

The relation for the gas compression ratio, $r$, across the shock jump can be derived from the Rankine-Hugoniot
condition including the pressure and energy flux of waves, and is given as the following cubic equation,
\begin{eqnarray}
b^2 M_{\rm A}^2 r \{ (\gamma-1)r^2 + [ M_{\rm A}^2(2-\gamma) - (\gamma+1)] r + \gamma M_{\rm A}^2 \}  \nonumber\\
+(M_{\rm A}^2-r)^2 \{ 2r\beta- M_{\rm A}^2 [\gamma+1-(\gamma-1) r ] \} =0,
\label{eqcomp}
\end{eqnarray}
for a given set of parameters, $M_{\rm s}$, $\beta$, and $b$ (VS99). Here, $\gamma=5/3$ is used for the ICM gas.

The bottom-left panel of Figure 3 shows the solution of Equation (\ref{eqcomp}), $r$,
for three beta's ($\beta=1$, 10, and 80) and $b=0.1$
in the Mach number range of $M_{\rm s} \lesssim 5$.
Since the background magnetic field is parallel to the shock flow (\ie parallel shocks) and
the transverse components of wave fields are small ($\delta B=0.1 B_0$), 
$r$ is almost identical to the gas compression ratio of gasdynamic shocks, 
$r_{\rm gas}=(\gamma+1)M_s^2/\{(\gamma-1)M_s^2+2\}$, regardless of $\beta$.
In fact, $r$ would deviate from $r_{\rm gas}$, 
{  only if $b$ is substantially large or $\beta$ is small.
In the same panel, two such cases with ($b=0.3$ \& $\beta=1$) and ($b=0.1$ \& $\beta=0.5$) are shown for comparison,
with the green and magenta lines, respectively, to illustrate such dependence.}

Following VS99, the {\it cross-helicity} is defined as
\begin{equation}
H_{\rm c} =  {{(\delta B^{\rm f})^2-(\delta B^{\rm b})^2} \over {(\delta B^{\rm f})^2+ (\delta B^{\rm b})^2}},
\label{eqhel}
\end{equation}
where $\delta B^{\rm b}$ and $\delta B^{\rm f}$ are the magnetic fields of backward and forward waves, respectively.
In the preshock region, backward waves are expected to be dominant for CR-mediated shocks (see Introduction),
so we assume $H_{\rm c1} \approx -1$.

For power-law energy spectra of waves with slope $q$, $I(k)\propto k^{-q}$,
the transmission and reflection coefficients for backward and forward waves, respectively, in the postshock region
are derived from the equations for transverse momentum and tangential electric field, as follows,
\begin{equation}
T\equiv {{\delta B_2^{\rm b}}\over {\delta B_1}} 
= {{r^{1/2}+1} \over {2r^{1/2}}} \left( r {{M_{\rm A}+H_{c1}} \over {M_{\rm A}+r^{1/2} H_{c1}} }\right)^{(q+1)/2},
\label{eqtran}
\end{equation}
\begin{equation}
R\equiv {{\delta B_2^{\rm f}}\over {\delta B_1}} 
= {{r^{1/2}-1} \over {2r^{1/2}}} \left( r {{M_{\rm A}+H_{c1}} \over {M_{\rm A}-r^{1/2} H_{c1}} }\right)^{(q+1)/2}
\label{eqrefl}
\end{equation}
\citep{vs98}.
Note that these coefficients are independent of the wavenumber.
According to hybrid simulations of collisionless shocks by \citet{caprioli14b}, for shocks with $M_{\rm A}\lesssim 30$
where resonant streaming instability dominantly operates, 
the spectrum of excited magnetic turbulence in the precursor is consistent with $I(k)\propto k^{-1}$.
So we adopt $q=1$.
With these coefficients, the downstream cross-helicity can be estimated as
\begin{equation}
H_{\rm c2} = H_{\rm c1} \cdot {{T^2-R^2}\over {T^2+R^2}}.
\label{eqdownhel}
\end{equation}

The top panels of Figure 3 show $T$, $R$, and $H_{\rm c2}$, calculated with $b=0.1$, $q=1$, and $H_{c1}=-1.$
One can see that incident backward waves are amplified across the shock with $T>1$,
while forward waves are generated with $0<R<1$
(greater $R$ for higher $\beta$) in the postshock region.
The ensuing downstream cross-helicity ranges $-1< H_{\rm c2} \lesssim -0.85$ for the shocks considered here. 
We note that the quasi-linear treatment adopted here should break down for non-linear waves, which are
expected to develop via streaming instabilities at strong shocks.

\section{Effects of Alfv\'enic Drift on DSA}

\subsection{Scattering Center Compression Ratio and CR Spectral Index}

\begin{figure*}[t!]
\vskip -0.2cm
\centering
\includegraphics[trim=2mm 2mm 2mm 2mm, clip, width=160mm]{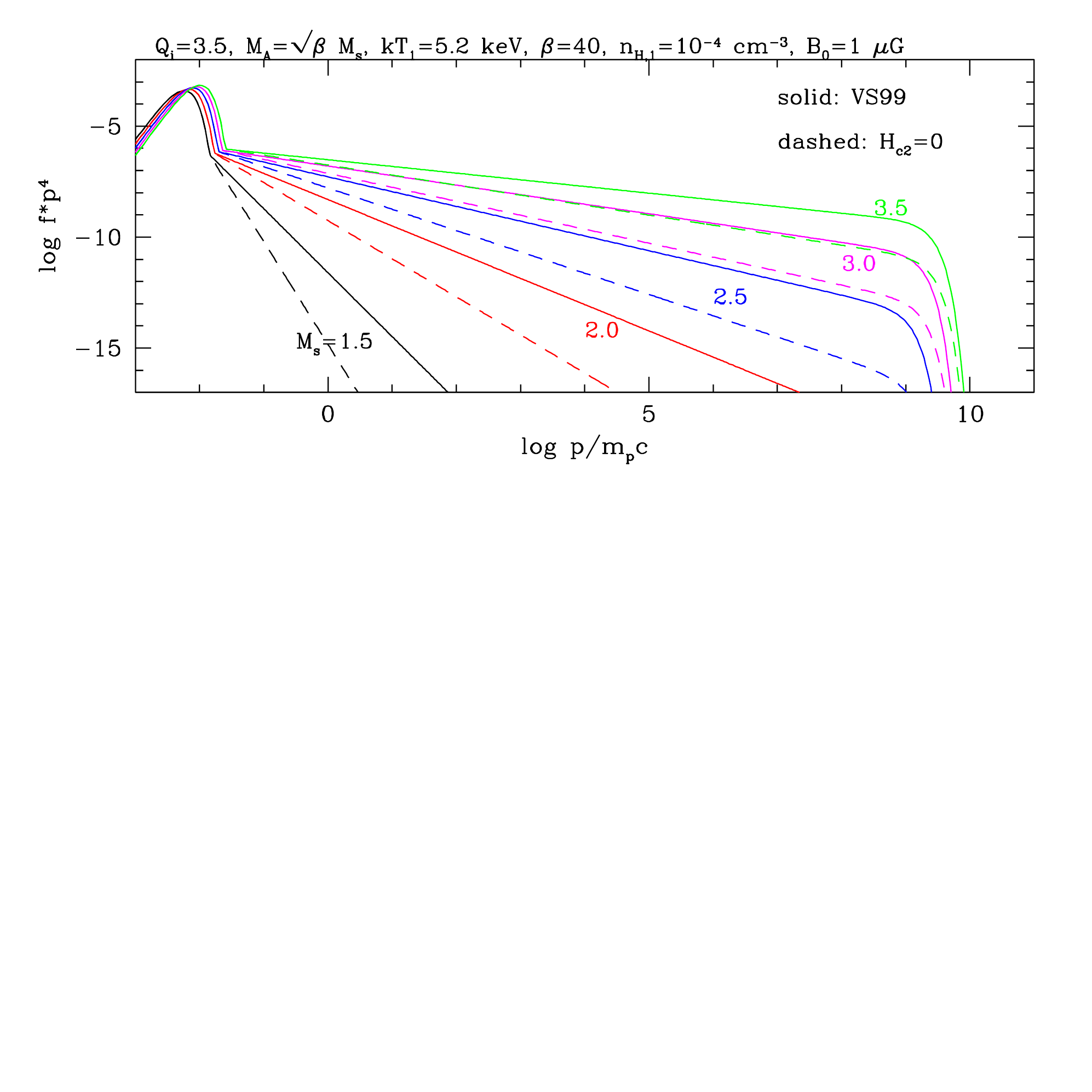}
\vskip -9.4cm
\caption{Test particle spectrum, $f(p)*p^4$, given in Equation (\ref{finj}), for models with different $M_{\rm s}$'s.
The model parameters are $Q_{\rm i}=3.5$, $k_{\rm B}T_1= 5.2 {\rm keV} $, $n_{\rm H1}=10^{-4} \cm3$, and $B_0=1 \muG $
($\beta = 40$). 
Each curve is labeled with $M_{\rm s}$.
The slopes of the power-law CR proton distributions, anchored to the postshock Maxwellian distributions, are
calculated with Equations (\ref{eqscatt}) and (\ref{eqgamma}) for 1D planar shocks.
The solid lines represent the models with $H_{\rm c2}$ estimated according to VS99, while the dashed lines show the
models with $H_{\rm c2}=0$ (isotropic waves).
}
\end{figure*}

The CR transport at shocks can be described by the diffusion-convection equation,
\begin{eqnarray}
{\partial f}\over \partial t} + (u+u_{\rm w}) {\partial f \over \partial x}
= {1\over{3}} {{\partial ( u+u_{\rm w}) }\over {\partial x}} \cdot p {{\partial f \over {\partial p}} \nonumber\\
+ {\partial \over \partial x} \left[\kappa(x,p){\partial f \over \partial x} \right]
+ {1\over{p^2}} {\partial \over \partial p}\left( p^2 D_{pp} {\partial f \over \partial p} \right),
\label{diffcon}
\end{eqnarray}
where $f(x,p,t)$ is the pitch-angle-averaged phase space distribution function
for CRs, $u$ is the flow speed, $u_{\rm w}$ is the local speed of scattering centers,
$\kappa(x,p)$ is the spatial diffusion coefficient, $D_{pp}$ is the momentum diffusion coefficient
\citep{skill75,bell78, schlickeiser89}.
The effects of Alfv\'enic drift enter through $u_{\rm w}$, which is here given as $u_{\rm w1} = H_{\rm c1} V_{\rm A1}$
and $u_{\rm w2}= H_{\rm c2} V_{\rm A2}$ in the preshock and postshock regions, respectively.

CR particles then experience the velocity change from $ u_1+ H_{\rm c1} V_{\rm A1}$ to
$u_2+ H_{\rm c2} V_{\rm A2}$ across the shock, since they are isopropized in the local wave frame.
Then the {\it compression ratio of scattering centers}, defined as the velocity jump of scattering centers,
is given as
\begin{equation}
r_{\rm sc} \equiv {{u_1+ H_{\rm c1} V_{\rm A1}}\over {u_2+ H_{\rm c2} V_{\rm A2}}
}=r {{M_{\rm A}+H_{c1}}\over {M_{\rm A}+r^{1/2} H_{c2}}}.
\label{eqscatt}
\end{equation}
Thus, $r_{\rm sc}$ can be different from the gas compression ratio, $r$, from Equation (\ref{eqcomp}), depending on the cross-helicity.
The bottom-middle panel of Figure 3 shows that $r_{\rm sc}$, calculated for 1D planar shocks (VS99), depends on $\beta$ and can be greater than $r_{\rm gas}$. 
But for $\beta \gg 1$, $r_{\rm sc}\approx r_{\rm gas}$, since $M_{\rm A} \gg 1$.

At weak cluster shocks, the CR pressure is dynamically insignificant,
that is, shocks are in the {\it test-particle regime}, in which the CR energy spectrum, $N(E)$, is
represented by a power-law form.
Then, its power-law index, $\Gamma$, is determined by $r_{\rm sc}$ as
\begin{equation}
\Gamma = {{r_{\rm sc} + 2}\over {r_{\rm sc} -1}}
\label{eqgamma}
\end{equation}
\citep{bell78}.
The bottom-right panel of Figure 3 shows $\Gamma$ calculated for 1D planar shocks.
Flattening of $N(E)$ due to Alfv\'enic drift could be substantial for $\beta \approx 1$ (red solid lines), 
for which even $\Gamma<2$ is predicted.
It can be seen that for $\beta \gg 1$ (see blue dot-dashed lines), $\Gamma \approx \Gamma_{\rm gas}$
since $r_{\rm sc}\approx r_{\rm gas}$.

\subsection{CR Acceleration Efficiency}

\begin{figure*}[t!]
\vskip -0.5cm
\centering
\includegraphics[width=180mm]{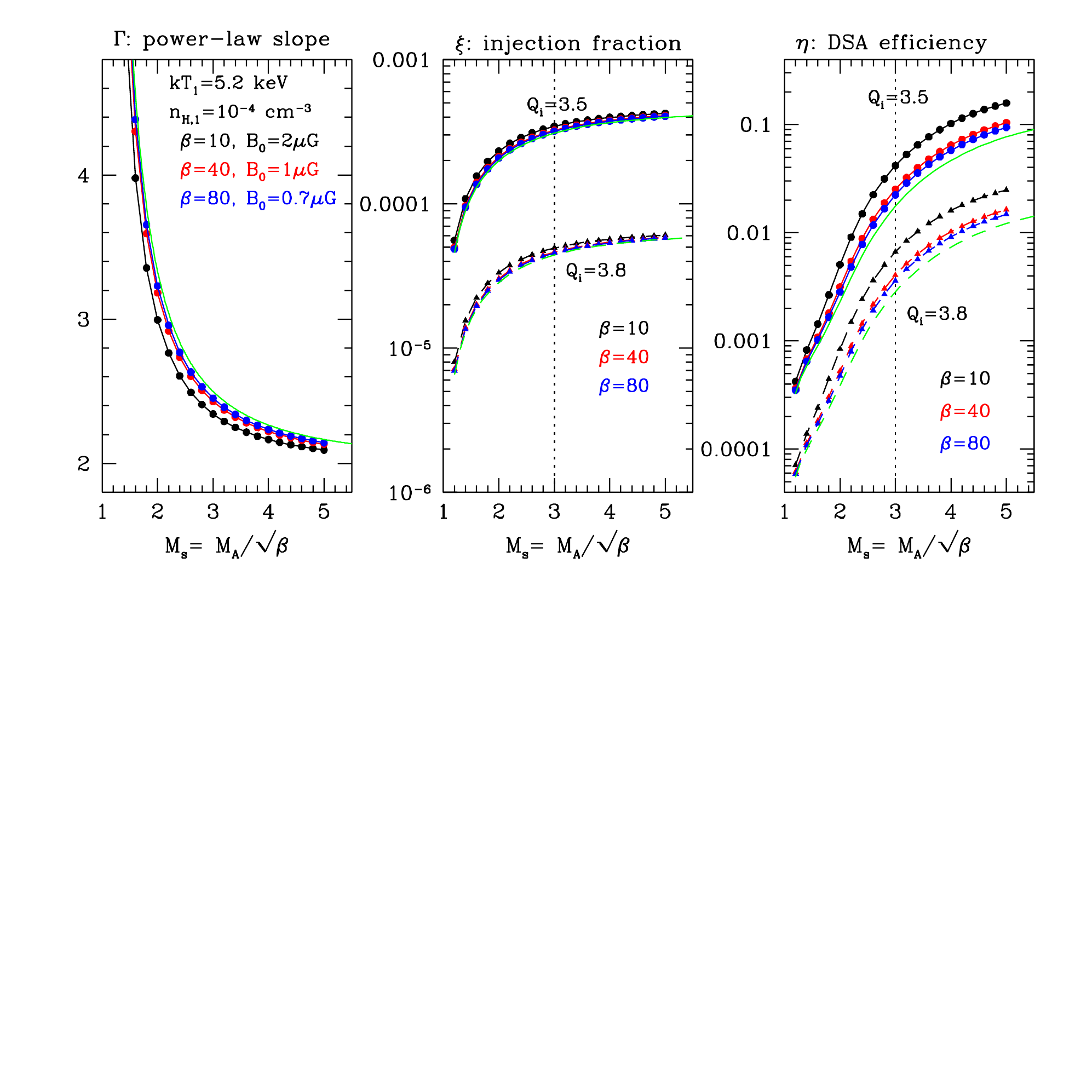}
\vskip -8.8cm
\caption{Left: Power-law slope, $\Gamma$, calculated with the scattering center compression ratio, $r_{\rm sc}$,
in Equation (\ref{eqscatt}).
The model parameters are $n_{\rm H1}=10^{-4} \cm3$, $k_{\rm B}T_1=5.2~{\rm keV}$, and $B_0=1 \muG (\beta/40)^{-1/2}$.
The value of beta is $\beta=10$ (black), 40 (red), and 80 (blue).
The postshock cross-helicity, $H_{\rm c2}$, is calculated by following VS99.
Middle: CR injection fraction, $\xi$, with $Q_{\rm i}=3.5$ (solid lines with circles),
and with $Q_{\rm i}=3.8$ (dashed lines with triangles).
Right: CR proton acceleration efficiency, $\eta$, calculated with the test-particle spectrum in Equation (\ref{finj})
with $Q_{\rm i}=3.5$ (solid lines with circles) and 3.8 (dashed lines with triangles).
The green lines show $\Gamma_{\rm gas}$ and $\eta$ for gasdynamic shocks without Alf\'enic drift.
}
\end{figure*}

In the test-particle regime, the amplitude of the CR proton spectrum can be fixed by setting it at the {\it injection momentum}, 
$p_{\rm inj}$, and then the momentum distribution function at the shock position, $x_s$, is given as
\begin{equation}
f(x_s,p)= f_N \left({p \over p_{\rm inj}}\right)^{-(\Gamma+2)} \exp \left[-\left({p \over p_{\rm cut}}\right)^2\right],
\label{finj}
\end{equation}
where $f_N$ is the normalization factor \citep{kang10}.
The cutoff momentum, $p_{\rm cut}$, represents the maximum momentum of CR protons that can be accelerated within the shock age,
$t_{\rm age}$, and is given as $p_{\rm cut} \propto u_1^2 B_0 t_{\rm age}$.
As long as $p_{\rm cut} \gg m_pc$, the CR energy density does not depend on its exact value if $\Gamma > 2$.

Here, we define $p_{\rm inj}$ as the minimum momentum above which protons can cross the shock transition
and participate in the Fermi I acceleration process, and describe it with the {\it injection parameter}, $Q_{\rm i}$, as
\begin{equation}
p_{\rm inj}\equiv Q_{\rm i} \cdot p_{\rm th},
\label{eqinj}
\end{equation}
where $p_{\rm th}=\sqrt{2m_pk_{\rm B}T_2}$ is the proton thermal peak momentum of the postshock gas with temperature $T_2$ \citep{kang10}. 
Using hybrid simulations, \citet{caprioli15} demonstrated that the injection momentum increases with the shock obliquity angle,
$\Theta_{\rm Bn}$, 
and {  $Q_{\rm i} \approx 3.3-4.6$} for quasi-parallel shocks ($\Theta_{\rm Bn}\lesssim 45^{\circ}$) with $M_{\rm A}=5-50$ and $\beta\approx1$.
{  The injection parameter should be affected by the strength of self-generated MHD turbulence, 
which in turn depends on $M_{\rm A}$ and $\beta$,
in addition to $\Theta_{\rm Bn}$. It is also expected to increase in time as the particle spectrum extends to higher energies
for strong shocks with $p^{-4}$ momentum distribution,
since the CR conversion efficiency cannot be greater than 100~\%.
More accurate estimation of $Q_{\rm i}$ for weak cluster shocks in high $\beta$ ICM plasmas, however, could be made only through kinetic plasma simulations, but its value has not yet been precisely defined \citep[see, e.g.,][]{caprioli14a,caprioli15}.}

Assuming that $f(x_s,p)$ is anchored to the postshock Maxwellian distribution at $p_{\rm inj}$,
the normalization factor is given as
\begin{equation}
f_N= {n_{\rm H2} \over \pi^{1.5} } p_{\rm th}^{-3} \exp(-Q_{\rm i}^2),
\label{eqnorm}
\end{equation}
where $n_{\rm H2}$ is the postshock hydrogen number density \citep{kang10}. 

Figure 4 illustrates how the test-particle spectrum in Equation (\ref{finj}) depends on the sonic Mach number, $M_{\rm s}$.
We adopt the relevant parameters for cluster shocks, $k_{\rm B}T_1=5.2$ keV, $n_{\rm H1}=10^{-4} \cm3$, and $B_0=1 \muG $,
resulting in $\beta\approx 40$.
{  We set $Q_{\rm i}=3.5$ as a representative value, since we here model mostly parallel shocks with small obliquity angles.
(Below, we also consider $Q_{\rm i}=3.8$ as a comparison case.)}
The CR spectra shown have the power-law indices, $\Gamma$'s, from Equations (\ref{eqscatt}) and (\ref{eqgamma}),
which are calculated with $H_{\rm c2}$ estimated according to VS99 for 1D planar shocks
(also with $H_{\rm c2}=0$, see Section 3.3).
The cutoff momentum for $t_{\rm age}=10^8$ yr is drawn for an illustrative purpose.

\begin{figure*}[t!]
\vskip -0.5cm
\centering
\includegraphics[width=180mm]{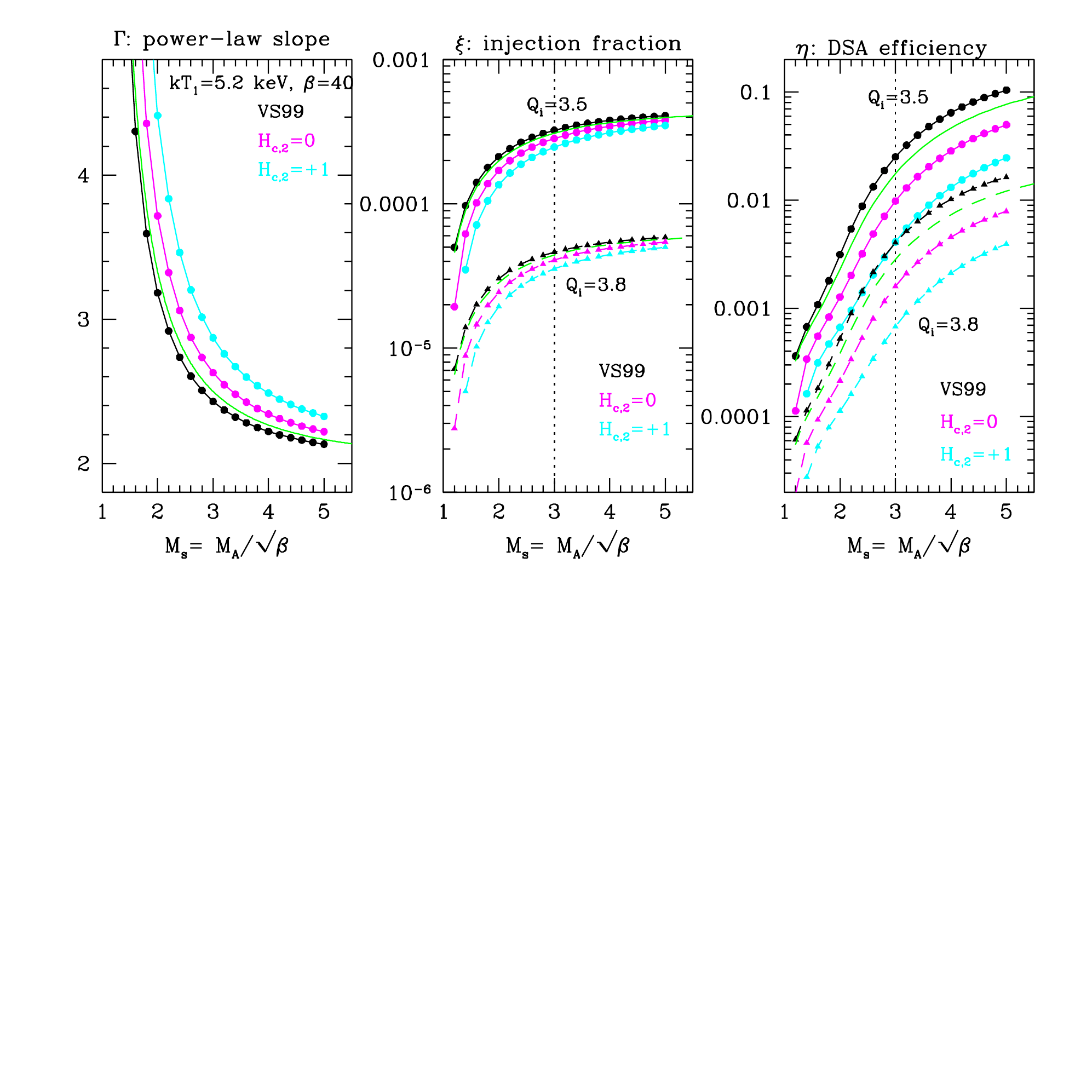}
\vskip -8.8cm
\caption{Left: Power-law slope, $\Gamma$, calculated with the scattering center compression ratio, $r_{\rm sc}$,
in Equation (\ref{eqscatt}).
The magenta and cyan lines are for $H_{\rm c2}=0$ and $H_{\rm c2}=+1$, respectively,
while black line shows the case with $H_{\rm c2}$ calculated by following VS99.
The model parameters are $n_{\rm H1}=10^{-4} \cm3$, $k_{\rm B}T_1=5.2$~keV, and $B_0=1~\muG$ ($\beta=40$).
Middle: CR injection fraction, $\xi$, with $Q_{\rm i}=3.5$ (solid lines with circles) and 3.8 (dashed lines with triangles). 
Right: CR acceleration efficiency, $\eta$, calculated with the test-particle spectrum given in Equation (\ref{finj})
with $Q_{\rm i}=3.5$ (solid lines with circles) and 3.8 (dashed lines with triangles). 
The green lines show $\Gamma_{\rm gas}$ and $\eta$ for gasdynamic shocks without Alf\'enic drift.
}
\end{figure*}

With the spectrum in Equation (\ref{finj}) {  and $\Gamma > 2$ for weak shocks}, the {\it CR injection fraction} can be estimated as
\begin{eqnarray}
\xi \equiv {1 \over n_{\rm H2}} \int_{p_{\rm min}}^{p_{\rm cut}}  4\pi f(r_{\rm sc},p) p^2 dp \nonumber\\
\approx {4 \over \sqrt{\pi}(\Gamma-1) } Q_{\rm i}^3 \exp(-Q_{\rm i}^2),
\label{eqinject}
\end{eqnarray}
{  if we take $p_{\rm min}=p_{\rm inj}$ as the lower boundary of the CR momentum distribution \citep{kang10}.
According to this definition, the CR injection fraction depends mainly on $\Gamma$ and $Q_{\rm i}$, since normally $p_{\rm cut} \gg m_pc$.}

We also define the {\it CR acceleration efficiency} as the ratio of the downstream CR energy flux 
to the shock kinetic energy flux, as follows,
\begin{equation}
\eta \equiv { f_{\rm CR} \over f_{\rm kin} }
={ {u_2 E_{\rm CR}} \over {(1/2) \rho_1 u_1^3}}
\label{eqeff}
\end{equation}
\citep{kang13}.
Here the postshock CR energy density is given as
\begin{equation}
E_{\rm CR}= 4\pi m_p c^2 \int_{p_{\rm min}}^{p_{\rm cut}} (\sqrt{ p^2 + 1} - 1)  f(x_s,p)  p^2 dp,
\label{eqcre}
\end{equation}
where the particle momentum $p$ is expressed in units of $m_pc$.
{  Again, we take $p_{\rm min}=p_{\rm inj}$ in the calculation of $E_{\rm CR}$ below.
Note that in general, the CR injection fraction 
and the DSA efficiency sensitively depend on how one specifies $p_{\rm min}$, 
since the CR number is dominated by nonrelativistic particles with $p \sim p_{\rm inj}$.
}

{  The left panel of Figure 5 shows the power-law slope, $\Gamma$, estimated with $H_{\rm c2}$, which is
calculated according to VS99. 
Here $\beta$ varies in the ranges relevant
to cluster shocks, $\beta=10-80$, so does the background magnetic field as $B_0= 1 \muG (\beta/40)^{-1/2}$.
One can see that at weak cluster shocks, $H_{\rm c2}$ based on VS99 could flatten the CR spectrum slightly, 
compared to gasdynamic shocks without Alf\'enic drift (green line).
But for $\beta \gg 1$ the dependence of $\Gamma$ on $\beta$ is rather weak.}

{  The middle and right panels of Figure 5 show the injection fraction, $\xi$, and 
the CR acceleration efficiency, $\eta$, respectively, calculated with the test-particle spectrum in Equation (\ref{finj})
with the slope $\Gamma$ shown in the left panel.
Here, the adopted values of $k_{\rm B}T_1$ and $n_{\rm H1}$ are the same as in Figure 4.
Both $\xi$ and $\eta$ strongly depend on $Q_{\rm i}$ through the normalization factor $f_N$,
due to the exponential nature of the tail in the Maxwellian distribution.
While $\xi \propto Q_{\rm i}^3 \exp(-Q_{\rm i}^2)$ from Equation (13),
the CR acceleration efficiency can be approximated as $\eta \propto Q_{\rm i}^5 \exp(-Q_{\rm i}^2)$
for weak shocks with power-law spectra much steeper than $p^{-4}$ (dominated by nonrelativisitc particles).
So $\xi$ decreases by a factor of 7 as $Q_{\rm i}$ increases from 3.5 to 3.8, while $\eta$ decreases roughly by a factor of 6 or so.}

For cluster shocks with $M_{\rm s}\lesssim 3$, $\xi \lesssim 3.2\times 10^{-4}$ and $\eta \lesssim 2.2\times 10^{-2}$ for $Q_{\rm i}=3.5$,
while $\xi \lesssim 4.6\times 10^{-5}$ and $\eta \lesssim 3.6\times 10^{-3}$ for $Q_{\rm i}=3.8$.
This indicates that the estimated CR injection fraction and acceleration efficiency could easily differ by an order of magnitude,
depending on the adopted $Q_{\rm i}$.
{  For parallel shocks with small obliquity angles (i.e., $\Theta_{\rm Bn}\lesssim 15^{\circ}$), 
however, we expect that $Q_{\rm i}$ is unlikely to be much larger than 3.8 \citep{caprioli15}.}

\subsection{Cases with $H_{\rm c2}\approx 0$ and $H_{\rm c2}\approx +1$}

The overall morphology of cluster shocks, induced mainly by merger-driven activities in turbulent ICMs,
is expected to be quite complex and different from simple 1D planar shocks \citep[see, e.g.,][]{vazza17,ha17}. 
Rather, it can be characterized by portions of spherically expanding shells, composed of multiple shocks
with different properties. 
In addition, vorticity is generated behind curved shock surfaces, leading to turbulent cascade over a wide range
of length scales and turbulent amplification of magnetic fields in the postshock flow \citep[see, e.g.][]{ryu08,vazza17}.
Then, downstream waves could be isotropized through various MHD and plasma processes in the postshock region, 
resulting in zero cross-helicity, $H_{\rm c2}\approx 0$ (equal strengths of $T$ and $R$).
Note that Fermi II acceleration should be operative in this case, but it is expected to be much less efficient 
than Fermi I acceleration.

In addition, as mentioned in the Introduction, the CR particle distribution peaks at the shock (\ie decreases downstream) in
spherical shocks or even in evolving planar shocks in which the CR pressure at the shock is increasing with time.
In that case, the gradient of $P_{\rm CR}$ is expected to damp backward waves, leaving dominantly forward waves 
with $H_{\rm c2} \approx +1$ in the postshock region \citep{bell78,zp08,caprioli09}.
Hence, we here quantitatively examine the effects of Alfv\'enic drift in these physically motivated cases
with $H_{\rm c2} = 0$ and $H_{\rm c2} = +1$, as phenomenological models.

In the panels for $r_{\rm sc}$ and $\Gamma$ of Figure 3, the magenta and cyan lines 
show $H_{\rm c2} = 0$ and $H_{\rm c2} = +1$ cases, respectively.
In fact, the scattering center compression ratio is minimized for $H_{\rm c2}=+1$ (see Equation (\ref{eqscatt})).
So this represents the case with the greatest impact of Alfv\'enic drift (the largest $\Gamma$).
Moreover, Figure 4 compares the models with $H_{\rm c2}$ estimated according to VS99 and the models with $H_{\rm c2}=0$ 
(isotropic waves), demonstrating how the Alfv\'enic drift may affect the CR spectrum.

Figure 6 shows $\Gamma$, $\xi$, and $\eta$ for the cases with $H_{\rm c2}=0$ (magenta lines) and
$H_{\rm c2}=+1$ (cyan lines), for the model parameters relevant to cluster shocks and $Q_{\rm i}=3.5$ and 3.8. 
The case for 1D planar shocks with $\beta=40$, calculated by following the VS99 approach (black lines),
where $-1 \lesssim H_{\rm c2} \lesssim -0.85$, is also plotted for comparison.
Again the green lines show the results for gasdynamic shocks without Alfv\'enic drift.
The scattering center compression ratio, $r_{\rm sc}$, is smaller for larger $H_{\rm c2}$,
resulting in steeper $\Gamma$, hence, smaller $\xi$ and $\eta$.
For weak cluster shocks with $2\lesssim  M_{\rm s}\lesssim 3$ and isotropic downstream waves with $H_{\rm c2}=0$,
the CR acceleration efficiency is $10^{-3}\lesssim \eta \lesssim 10^{-2}$ for $Q_{\rm i}=3.5$ and
$2\times 10^{-4}\lesssim \eta \lesssim 1.5\times 10^{-3}$ for $Q_{\rm i}=3.8$.
For the case of dominantly forward waves with $H_{\rm c2}=+1$, on the other hand,
$7\times 10^{-4}\lesssim \eta \lesssim 4\times 10^{-3}$ for $Q_{\rm i}=3.5$ and
$10^{-4}\lesssim\eta \lesssim 7\times 10^{-4}$ for $Q_{\rm i}=3.8$.
Our results indicate that $\eta$ could be reduced by ``a factor of up to $\sim 5$'' due to Alfv\'enic drift alone.
Thus, to quantify the CR acceleration efficiency, it could be crucial not only to constrain the injection parameter
$Q_{\rm i}$ through plasma simulations, but also to account for Alfv\'enic drift effects.

\section{Summary}

We study the effects of Alfv\'enic drift on the DSA of CR protons at weak shocks in high beta ICM plasmas.
We assume that upstream Alfv\'en waves are self-excited by CR protons via resonant streaming
instability at parallel shocks \citep{lucek00,schure12}.
Such waves are mostly backward, moving anti-parallel to the background flow \citep{bell78}, so they can be characterized by
the cross-helicity of $H_{\rm c1}\approx -1$ (see Equation (\ref{eqhel}) for the definition of $H_{\rm c}$).
Since CR protons are scattered and isotropized in the local wave frame, 
the scattering center compression ratio, $r_{\rm sc}$, in Equation (\ref{eqscatt}),
which accounts for the mean drift of Alv\'en waves, determines the spectral index, 
$\Gamma$, of the CR spectrum in the test-particle limit.

We first consider 1D planar shocks where the transport of Alfv\'en waves across the shock transition
is described in the small wave amplitude limit ($b \equiv \delta B/B \ll 1$) \citep[][and V99]{vs98}.
In this limit, as noted by VS99, Alfv\'enic drift may increase or decrease $r_{\rm sc}$, 
depending on the shock parameters. 
This results in the CR spectra either flatter or steeper, compared to that for gasdynamic shocks without Alfv\'enic drift.
For shocks with $M_{\rm s}\lesssim 3$ and $\beta \equiv ({M_{\rm A} / M_{\rm s}})^2 \sim 40-80$,
a mixture of backward and forward waves are present in the postshock region with the postshock cross-helicity 
estimated to $-1 \lesssim H_{\rm c2} \lesssim -0.85$, leading to only a slight decrease of $\Gamma$ (see Figure 3).
That is, for weak cluster shocks, $r_{\rm sc}\approx r_{\rm gas}$ and $\Gamma \approx \Gamma_{\rm gas}$, 
and so the effects of Alfv\'enic drift on the DSA efficiency are only marginal (see Figure 5).
 
We then consider two additional, physically motivated cases:
(1) downstream waves are isotropic with $H_{\rm c2}\approx 0$,
and (2) they are dominantly forward with $H_{\rm c2}\approx +1$.
The former could be realistic, if waves are isotropized via a variety of MHD and plasma
processes including turbulence while they cross the shock transition.
The latter may be relevant, if the CR pressure distribution peaks at the shock as in spherical SNR shocks
or evolving planar shocks. 
In these two cases, Alfv\'enic drift causes the CR spectrum to be steeper,
which results in significant reductions of the CR injection fraction, $\xi$,
and the CR acceleration efficiency, $\eta$ (see Figure 6).
In the case of $H_{\rm c2}\approx +1$, for example,
the CR proton acceleration efficiency
for shocks with $M_{\rm s} \lesssim 3$ and $\beta \approx 40$
could be reduced by ``a factor of up to 5'', compared to that for gasdynamic shocks.
So we conclude that the Alfv\'enic drift effects on the DSA efficiency could be substantial at weak cluster shocks.

We note that the CR acceleration efficiency is most sensitive to the injection momentum, or, the
injection parameter, $Q_{\rm i}$, defined in Equation (\ref{eqinj}).
Increasing $Q_{\rm i}$ from 3.5 to 3.8 (about 10 \%), for instance, reduces $\eta$ by a factor of $5-7$. 
{  For parallel shocks with small obliquity angles, we expect that $Q_{\rm i} = 3.5 - 3.8$ would be a reasonable range.}
Thus, in order to reliably estimate the CR proton acceleration efficiency at weak cluster shocks,
it is important to understand the kinetic plasma processes that govern the particle 
injection to Fermi I acceleration at collisionless shocks at high beta plasmas.

{  We suggest $\eta$ could vary in a wide range of $10^{-4}-10^{-2}$ for weak cluster shocks with $M_{\rm s}\approx2-3$,
depending on $H_{c2}$, $\Theta_{\rm Bn}$, and $\beta$. 
Such estimate could be smaller by up to an order of magnitude than that adopted in the previous studies such as \citet{vazza16}.
So this study implies that there remains room for the DSA prediction for CR proton acceleration at cluster shocks
to be compatible with non-detection of $\gamma$-ray emission from galaxy clusters \citep{ackermann14, ackermann16}.
Yet, we emphasize that eventually
detailed quantitative studies of DSA at weak cluster shocks using kinetic plasma simulations
should be crucial for solving this problem.}

\acknowledgements
{  We thank the anonymous referee for constructive comments.}
H.K. was supported by the Basic Science Research Program of the NRF of Korea through grant 2017R1D1A1A09000567.
D.R. was supported by the NRF of Korea through grants 2016R1A5A1013277 and 2017R1A2A1A05071429.
The authors also thank R. Schlickeiser for helpful comments during the initial stage of this work.

\end{document}